# Low Energy Surface Flashover for Initiation of Electric Propulsion Devices

Yunping Zhang, Omar Dary and Alexey Shashurin
*School of Aeronautics and Astronautics, Purdue University, West Lafayette, IN 47906, USA*

**Abstract**

An approach to utilize Low Energy Surface Flashover (LESF) for triggering the discharge in electric propulsion systems has been demonstrated. LESF uses conventional surface flashover mechanism with limited duration of high-current stage of the flashover below <100-200 ns. This eliminates the damage to the LESF assembly and allows robust operation of the same assembly for $>1.5 \cdot 10^6$ consecutive flashovers. The amount of the seed plasma created in the individual LESF event was demonstrated to be sufficient to trigger a moderate current arc which models a discharge in an electric propulsion system.

## 1. Introduction

Recently there has been a rapid increase of interest in small satellites, such as CubeSats, which are usually launched as secondary payloads and are useful as instruments of targeted investigations to augment the capabilities of large space missions and enable new kinds of measurements [1]. With an on-board propulsion system, CubeSats are able to achieve orbital maneuvers, formation flying, constellation maintenance and precise attitude control [1]. Chemical propulsion as one candidate for propelling smaller spacecraft into outer space has the advantage of large thrust but presents severe concerns due to its requirement for large propellant mass, high temperature and pressure, and a threat to the main payloads posed by the reactive propellant materials. Electric propulsion, in comparison, has very high exhaust velocity and fuel efficiency. Depending on the mechanism of acceleration, traditional electric propulsion systems are generally divided into three categories [2,3], Electrothermal [4,5], Electrostatic [6,7], and Electromagnetic [8,9,10]. While there are many different electric propulsion technologies for CubeSats currently in research/ testing, such as Pulsed Plasma Thruster [1], Miniature Xenon Ion Thruster [1,11], Electrospray Thruster [1,12-15] and Vacuum Arc Thruster [1,2,16-25], such propulsion systems are still at their infancy and mostly remain <7 in the Technology Readiness Level scale used by NASA [1].

One of the central part of electrical propulsion systems are ignitor subsystems, which are required for the discharge initiation. Generally, there are many different methods to ignite a discharge in vacuum, among which are initiation using gas injection, high voltage breakdown, mechanical actuators for drawn arcs, and fuse wire explosion [20,26,27], etc. Other methods such as the triggerless method use vaporization of conductive coating between the anode and cathode [28]. All these triggering mechanisms operate by providing seed plasma required to bridge the electrodes and initiate the discharge.

A robust and compact ignitor which can reliably trigger the discharge in the electrical propulsion system throughout the entire operational lifetime is challenging. While the triggering methods considered above are capable of discharge initiation, they have significant drawbacks from the prospective of propulsion applications. Indeed, the necessity to carry a gas storage tank for the gas injection triggering methods [26], and the need to utilize a high voltage source in high voltage breakdown techniques [20,27] are adding to weight and complexity of the ignitor. Use of the fuse wire explosion for the discharge ignition is highly unreliable if multiple ignitions are expected [27]. The triggerless approach suffers from electrode/film assembly damage after relatively low number of triggering events 1,000 – 10,000 [29].

One particular type of vacuum discharge triggering is a surface flashover. In the surface flashover, two electrodes are separated by an insulating layer and the breakdown over the insulating surface is initiated at application of high voltage that exceeds the breakdown threshold $V_{br}$. Studies of the surface flashover of insulators have been mainly driven by the high voltage vacuum devices' applications. High voltage holdoff capability is desired for these devices, and surface flashover and subsequent breakdown are undesirable effects. Thus, surface flashover was studied from the perspective of the ultimate goal to reduce the probability of the said events and increase the holdoff voltage capability of the device. It has been shown that surface flashover in vacuum depends upon many parameters, such as material, geometry, processing history of the insulator, the applied voltage waveform and its duration, and on the number of previous flashovers [30,31]. The breakdown voltage of insulators was found to be independent of pressure in the ranges of $5 \times 10^{-3}\ Torr\ to\ 10^{-7}\ Torr$ [32]. Typically, surface flashover can be broken down in a three-stage process [33]. It starts with stage 1 lasting for about 10 ns when electrons are emitted from Cathode Triple Junction (CTJ). Stage 2 (100-400 ns) is associated with Secondary Electron Emission Avalanche (SEEA) development since some electrons emitted from the CTJ impact the surface of the insulator and produce secondary electrons. On the stage 3 (>100-400 ns) desorption of gases from the insulator surface at SEEA development occurs, Townsend breakdown develops in these desorbed gases and high current arc (>10-100 Amperes) is establishing in the assembly [33-38].

Therefore, surface flashover was studied thoroughly by the high voltage vacuum devices' community. This classic flashover is associated with overheating of the flashover electrode assembly and permanent damage to the assembly after several flashover events due to high current arcs developing in the assembly on the stage 3 of the flashover event. However, a possibility to limit the energy of the surface flashover in order to shorten/eliminate the high-current stage 3 of the flashover event with an ultimate goal to reduce/eliminate the damage to the flashover electrode assembly and use it as an ignitor for the discharge in propulsion system was not studied earlier. In this work we have studied the novel approach to significantly reduce the energy of a surface flashover in order to achieve large number of flashovers (>1.5·10$^6$) with the same electrode assembly without significant damage or degradation to the assembly and demonstrate potential of using this approach for ignition of electric propulsion systems such as micro thrusters. The proposed approach is referred to as Low Energy Surface Flashover (LESF) in the following description.

## 2. Experimental details

The experiments were conducted in high vacuum setup pumped down to $5 \cdot 10^{-6} - 3 \cdot 10^{-5}$ Torr. Electrode assembly shown in Figure 1 was utilized in the experiments. It consists of a 0.635 mm thick non-porous alumina ceramic sheet clamped between the two 10 mm ×10 mm × 0.5 mm copper electrodes bonded to the ceramic sheet with a low vapor pressure epoxy. The electrodes were sanded with 600 grit sandpaper and the side on which the surface flashover events occur was additionally sanded using 514 grit diamond wheel.

This study has been intended to initiate the breakdown at potentially lower voltages. To this end, alumina ceramics was chosen as an insulator material since it is characterized by the relatively low surface flashover breakdown voltages about 5-10 kV/mm [39]. In addition, the insulator thickness was significantly reduced (down to <1 mm) in comparison to that normally used in surface flashover studies (>1 cm). While surface flashover in vacuum may require high voltages to discharge, this study was focused on attempts to limit the $V_{br}$ to the range of around 10-15 kV.

Two high voltage power supplies have been used to initiate the surface flashover in the experiments, namely Eagle Harbor Nanosecond Pulser model NSP-3300-20-F (<110ns, <20 kV) and Bertan Series 225-20R DC power supply (<20kV, <1mA). DC power supply Sorenson X60-28 (<60V, <28A) was used to support a main discharge triggered by the flashover (see details below). Electric characteristics of the discharges were measured by Tektronix P6015A passive high voltage probe, Pearson 2100 and Bergoz FCT-028-0.5-WB current monitors. Fast photographing of the surface flashover was captured by Princeton Instruments PI-MAX4 ICCD camera.

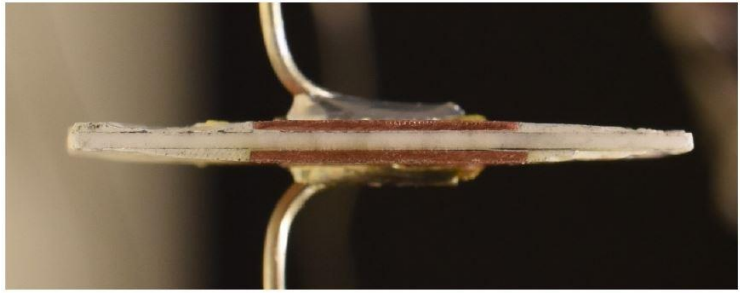

**Figure 1.** Photograph of the flashover assembly used in experiments after conducting >1.5·10⁶ flashover events.

### 3. Experimental results and discussions

Figure 2 shows evolution of the voltage required to breakdown the flashover electrode assembly ($V_{br}$) in >1.5·10⁶ flashover events ($N$). The flashover assembly shown in Figure 1 was used in these experiments. Eagle Harbor pulser was utilized with pulse amplitudes up to 15 kV and pulse duration of 110 ns in order to limit the duration of the high-current stage 3 of the flashover associated with high-current arcing [30,31]. These short flashover events with duration $\tau_{fl}\leq$100-200 ns are referred to as Low Energy Surface Flashover (LESF). Schematics of the utilized electrical circuit are shown in the insert of Figure 2. For the first 500 initial pulses the pulser was operated at single pulse mode, and then the pulse repetition rate was gradually increased from 1 to 200 Hz.

One can see from Figure 2 that the breakdown voltage increased rapidly from 2.9 kV during the first hundred pulses and then approached saturation in the range of 10-14 kV for $N$ >1000. The initial increase of $V_{br}$ is a well-known effect of conditioning of the insulator surface which is associated with removal of surface gas, removal of surface contaminants, or removal of emission sites [30,31]. One can see that $V_{br}$ continued to increase slowly in the following flashover events ($N$ >1000) and reached about 14.5 kV after about 10⁶ breakdowns. This nearly saturated region is associated with fully conditioned sample. Breakdown voltage was highly variable within 40% due to natural variability of the flashover events.

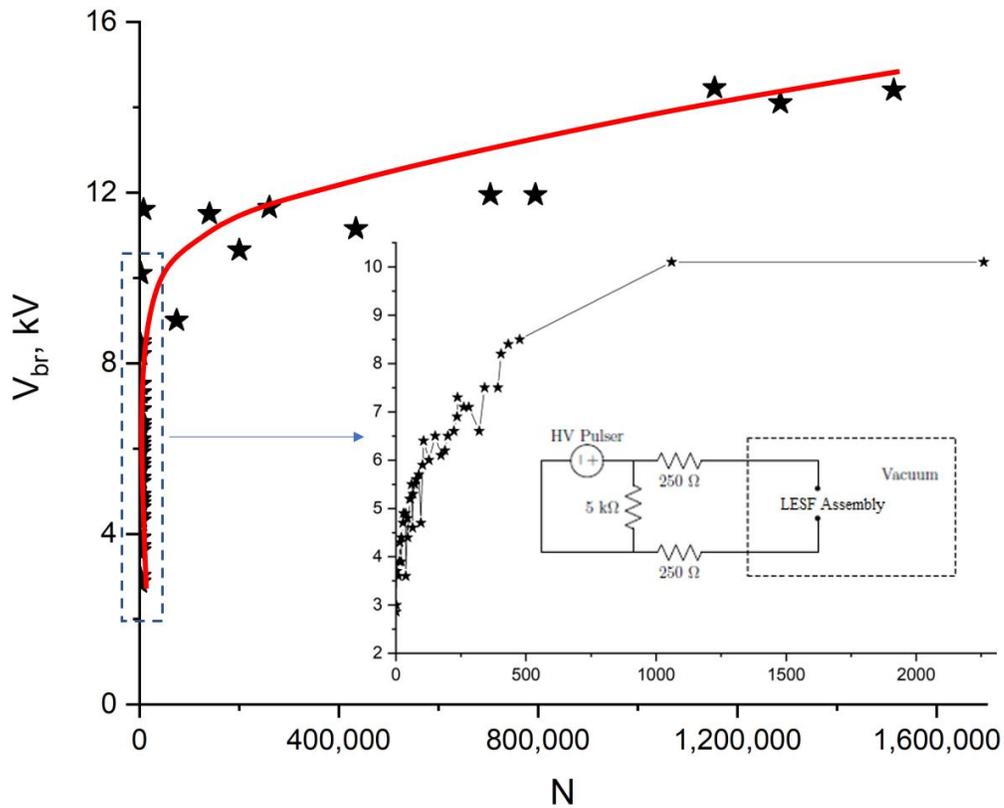

**Figure 2. Evolution of the breakdown voltage of the electrode assembly ($V_{br}$) over >1,500,000 flashover events ($N$).**

It is important to note, that the data presented in the Figure 2 was obtained using the same electrode assembly which was run without failure or damage to the electrodes/insulator for more than 1,500,000 pulses. The photograph shown in Figure 1 demonstrates that assembly after conducting >1.5·10⁶ flashover events. Only minor ablation of the electrode assembly can be visually observed at the electrode-insulator interface. This finding confirms that the approach of limiting the duration/energy of individual surface flashover event can ensure very long operational lifetime of the same electrode assembly.

An individual flashover event of a fully-conditioned assembly was studied in detail using the circuit shown schematically in Fig 3a. In this case, DC high voltage was applied to the electrodes through a current limiting resistor $R_{lim}$=100 kΩ connected in series to the electrodes using Bertan 225-20R DC power supply. Duration of the surface flashover ($\tau_{fl}$) was a free parameter in these experiments, and it is governed by the amount of the energy available in the capacitive circuit formed by the flashover assembly and the leads indicated by the dashed area in the Fig 3a (see details below). Simultaneous measurements of the electrical parameters of the flashover along with the series of photographs (3 ns exposure time) taken by the ICCD camera are presented in Fig. 3b-d.

One can see in Fig. 3b that initiation of the flashover was indicated by an instant drop of voltage (near $t≈0$) and start of high frequency current oscillations. These current oscillations are associated with the resonant ringing in the *LC*-circuit formed by the flashover assembly shortened by the plasma column (see dashed area in Figure 3a behind the current limiting resistor). Indeed, prior to the flashover the electrode assembly is equivalent to a capacitor ($C$) charged to the voltage ($V_{br}$) as shown in Figure 3c. This capacitance was measured prior to the experiment to be 7 pF and thus resulting in total energy stored in the capacitor: $\frac{CV_{br}^2}{2}$=0.07 mJ. Creation of the plasma in the flashover event causes immediate short of one side of the assembly by plasmas, while the other side of the assembly is nearly opened (see Figure 3c), since current replenishment through the large 100 kΩ resistor on the short flashover timescale ($\tau_{fl}$) is negligible ($R_{lim}C \gg \tau_{fl}$). Thus, total energy stored in the capacitor is oscillating between the open-ended capacitive side of the assembly and shortened by the plasmas inductive side of the assembly. The inductance of the shortened assembly is governed by the inductance of the leads and it was measured prior to experiments to be $L_w$=0.5 µH. One can see in the Figure 3b that period of the current oscillations was about 10-15 ns, which is consistent with the theoretical estimation for the resonant oscillation period in the *LC*-circuit: $2\pi\sqrt{L_w C}$=12 ns. Note, we do not present the evolution of the discharge voltage after the flashover start in Figure 3b, since these measurements are far beyond frequency bandwidth of the Tektronix P6015A probe (75 MHz) used in the experiments.

Oscillations of the discharge current were peaked at around 15 A around $t≈0$ and decayed on the time scale of about 50 ns according to the Figure 3b. This decay time provides an estimate of flashover duration time $\tau_{fl}$ as follows from the experimental results. Indeed, $\tau_{fl}$ was directly evaluated by means of fast photographing conducted by ICCD camera as shown in Figure 3d. These visual observations indicate that flashover decays on the approximately same time scale of about 50 ns. Note, even though driven by the DC high voltage source, the duration of the flashover presented in the Figure 3 was short $\tau_{fl}$< 100 ns and thus, it operated in the LESF mode.

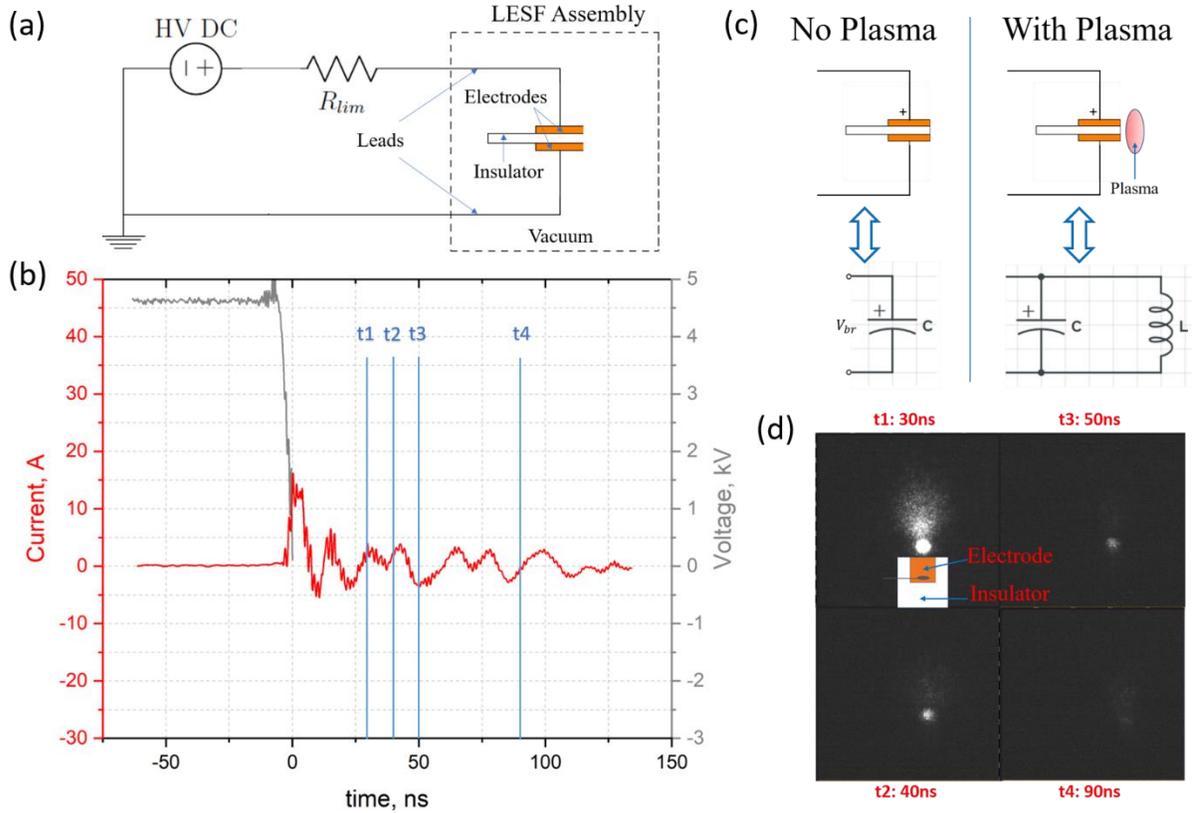

**Figure 3. Simultaneous measurements of the electric parameters and fast photographing of the surface flashover. a) Schematics of the electric circuitry used in the experiments. b) Voltage and current waveforms during the flashover event. Moments of time *t1-t4* correspond to the images shown in Figure 3d. c) Equivalent circuits of the LESF assembly and leads without and with plasma. d) Photographs of the LESF event taken in the moments of time *t1-t4* taken by the ICCD camera (exposure time=3ns).**

Duration of the flashover event $\tau_{fl}$ driven by the circuitry shown in the Figure 3a can be controlled by adjusting the amount of initial energy stored in the flashover/leads assembly prior to the event $E_0 = \frac{CV_{br}^2}{2}$ (see dashed area in the Figure 3a). Specifically, $\tau_{fl}$ can be increased if larger energy $E_0$ is used. To demonstrate this, an additional capacitor was inserted in parallel to the LESF assembly to increase the capacitance and energy stored in the circuit prior to the flashover event. The tests were conducted with two capacitances $C$ =7 and 100 pF and the corresponding initial energies stored in the assemblies were $E_0$ =0.35 and 5 mJ, respectively ($V_{br}$ was about 10 kV in both cases). Current waveforms for $E_0$=0.35 and 5 mJ are presented in Fig. 4. One can see that flashover duration $\tau_{fl}$ increased from about 50 ns to about 200 ns when initial energy $E_0$ increased from 0.35 to 5 mJ. In addition, the increase of capacitance to $C = 100$ pF led to the corresponding increase of the oscillations' period to about 50 ns which is in agreement with the theoretical estimation $2\pi\sqrt{L_w C}$=44 ns.

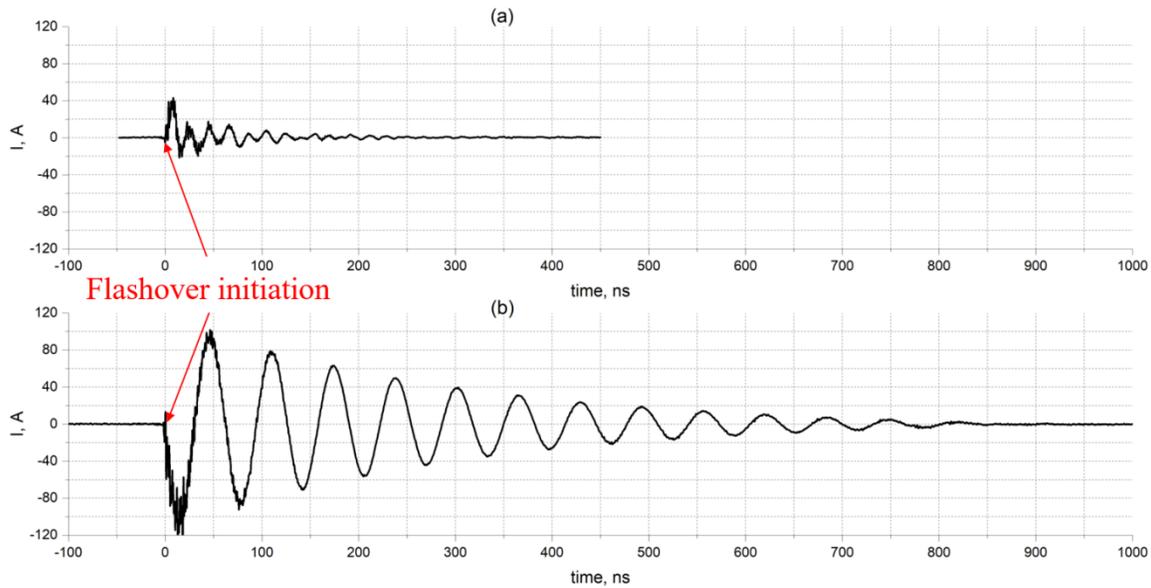

**Figure 4. Flashover current waveforms for two values of energy stored in the assembly just prior to the flashover event: (a) 0.35 mJ and (b) 5 mJ.**

In the following description we have evaluated whether the proposed approach can satisfy the purpose of triggering the discharge in the electric propulsion system. To this end, the LESF assembly was tested as an igniter for a moderate current vacuum arc and influence of initial energy $E_0$ on the ignition success was determined. Vacuum arc was used to model the discharge in the electric propulsion system in these experiments. An additional anode was placed at the distance $d$ from the LESF assembly as shown in Figure 5. The anode was biased to a voltage of +60 VDC with respect to cathode of the flashover assembly as shown in Figure 5a. The initial energy $E_0$ supplied to the flashover was varied by reducing the capacitance of the flashover/leads assembly.

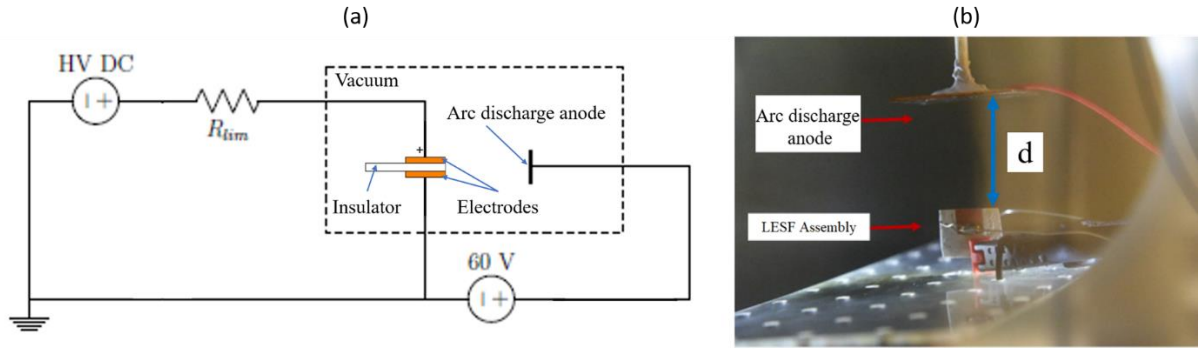

**Figure 5. Electric circuit used to test LESF as an ignitor of moderate current arc discharge.**

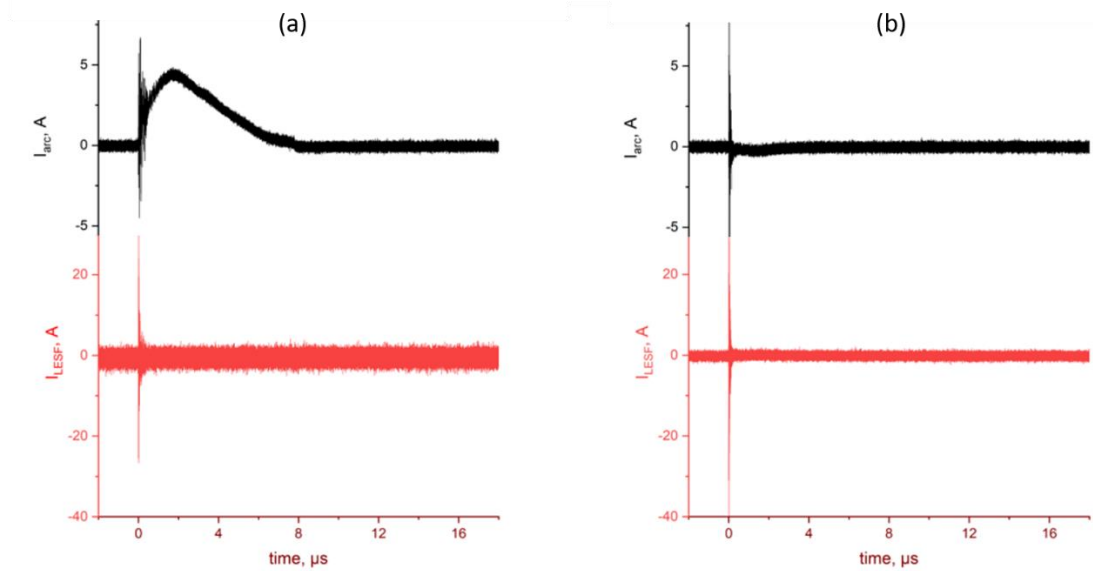

**Figure 6.** (a) Successful initiation of arc discharge by means of LESF ignitor with $E_0$ =0.73 mJ. (b) Unsuccessful attempt of ignition with $E_0$=0.39 mJ.

For $d$=4 cm, a successful ignition of the arc discharge was observed with the initial energy $E_0$=0.73 mJ in 16 out of 20 trials, while $E_0$=0.39 mJ failed to ignite the arc. The successful initiation of the arc discharge is demonstrated by the arc current pulse of about $I_{arc}$=5 A lasting for about 8 μs as shown in Figure 6a. For $d$=2 cm, a successful ignition of the arc discharge was observed in every try with the initial energy $E_0$=0.73 mJ, while $E_0$=0.39 mJ led to the ignition of 4 out of 20 tries. It can be inferred that seed plasma created by the LESF considered in this work is sufficient to trigger the arc discharge used to model the discharge in electric propulsion system. In addition, closer proximity of the arc anode to the flashover assembly facilitates the ignition due to higher density of the seed plasma in the gap.

## 4. Conclusions

Proposed here approach is paving the way for utilization of Low Energy Surface Flashover for triggering the electric propulsion systems. We have demonstrated that LESF electrode assembly can withstand extended operation for >1,500,000 breakdowns without significant damage to the electrode assembly. The amount of seed plasma produced in the single flashover event is sufficient to trigger a moderate current arc which was used to model a discharge in the electric propulsion system.

## 5.Reference

[1] Board, Space Studies, and National Academies of Sciences, Engineering, and Medicine. Achieving Science with CubeSats: Thinking Inside the Box. National Academies Press, 2016.

[2] Keidar, Michael, Taisen Zhuang, Alexey Shashurin, George Teel, Dereck Chiu, Joseph Lukas, Samudra Haque, and Lubos Brieda. "Electric propulsion for small satellites." Plasma Physics and Controlled Fusion 57, no. 1 (2015).

[3] Jahn, Robert G. Physics of electric propulsion. Courier Corporation, 2006.

[4] Howard, James M. "The resistojet." ARS (Am. Rocket Soc.) J.32 (1962).

[5] John, R. R., S. Bennett, and J. F. Connors. "Arcjet engine performance: experiment and theory." AIAA (Am. Inst. Aeron. Astronaut.) J. 1 (1963).

[6] H. R. Kaufman. Electric propulsion for spacecraft. New Science, 23:263, 1964.

[7] Goebel, Dan M. "Ion source discharge performance and stability." The Physics of Fluids 25, no. 6 (1982): 1093-1102.

[8] Jahn, Robert G., and Edgar Y. Choueiri. "Encyclopedia of Physical Science and Technology." Academic Press 3 (2002): 125-141.

[9] Cann, Gordon L., and Gary L. Marlotte. "Hall current plasma accelerator." AIAA Journal 2, no. 7 (1964): 1234-1241.

[10] Ahedo, Eduardo. "Plasmas for space propulsion." Plasma Physics and Controlled Fusion 53, no. 12 (2011): 124037.

[11] Conversano, Ryan W., and Richard E. Wirz. "Mission capability assessment of cubeSats using a miniature ion thruster." Journal of Spacecraft and Rockets 50, no. 5 (2013): 1035-1046.

[12] Lozano, Paulo, Manuel Martínez-Sánchez, and Jose M. Lopez-Urdiales. "Electrospray emission from nonwetting flat dielectric surfaces." Journal of Colloid and Interface Science276, no. 2 (2004): 392-399.

[13] Velásquez-García, Luis Fernando, Akintunde Ibitayo Akinwande, and Manuel Martinez-Sanchez. "A planar array of micro-fabricated electrospray emitters for thruster applications." Journal of Microelectromechanical Systems 15, no. 5 (2006): 1272-1280.

[14] Krejci, David, Fernando Mier-Hicks, Corey Fucetola, Paulo Lozano, Andrea Hsu Schouten, and Francois Martel. "Design and Characterization of a Scalable ion Electrospray Propulsion System." (2015).

[15] Krejci, David, Fernando Mier-Hicks, Robert Thomas, Thomas Haag, and Paulo Lozano. "Emission characteristics of passively fed electrospray microthrusters with propellant reservoirs." Journal of Spacecraft and Rockets 54, no. 2 (2017): 447-458.

[16] Polk, James E., Michael J. Sekerak, John K. Ziemer, Jochen Schein, Niansheng Qi, and Andre Anders. "A theoretical analysis of vacuum arc thruster and vacuum arc ion thruster performance." IEEE Transactions on Plasma Science 36, no. 5 (2008): 2167-2179.

[17] Keidar, Michael, Jochen Schein, Kristi Wilson, Andrew Gerhan, Michael Au, Benjamin Tang, Luke Idzkowski, Mahadevan Krishnan, and Isak I. Beilis. "Magnetically enhanced vacuum arc thruster." Plasma Sources Science and Technology 14, no. 4 (2005): 661.

[18] Zhuang, Taisen, Alexey Shashurin, Thomas Denz, Michael Keidar, Patrick Vail, and Anthony Pancotti. "Performance characteristics of micro-cathode arc thruster." Journal of Propulsion and Power 30, no. 1 (2014): 29-34.

[19] Lukas, Joseph, George Teel, Jonathan Kolbeck, and Michael Keidar. "High thrust-to-power ratio micro-cathode arc thruster." AIP Advances 6, no. 2 (2016): 025311.

[20] Boxman, Raymond L., David M. Sanders, and Philip J. Martin, eds. Handbook of vacuum arc science & technology: fundamentals and applications. William Andrew, 1996.

[21] Juttner, B. "The dynamics of arc cathode spots in vacuum." Journal of Physics D: Applied Physics 28, no. 3 (1995): 516.

[22] Beilis, Isak I. "State of the theory of vacuum arcs." IEEE transactions on Plasma Science 29, no. 5 (2001): 657-670.

[23] Daalder, J. E. "Erosion and the origin of charged and neutral species in vacuum arcs." Journal of Physics D: Applied Physics 8, no. 14 (1975): 1647.

[24] Anders, André, and George Yu Yushkov. "Ion flux from vacuum arc cathode spots in the absence and presence of a magnetic field." Journal of Applied Physics 91, no. 8 (2002): 4824-4832.

[25] Kimblin, C. W. "Erosion and ionization in the cathode spot regions of vacuum arcs." Journal of Applied Physics 44, no. 7 (1973): 3074-3081.

[26] Lafferty, J. M. "Triggered vacuum gaps." Proceedings of the IEEE 54, no. 1 (1966): 23-32.

[27] Farrall, George A. "Arc Ignition." In Handbook of Vacuum Arc Science and Technology, pp. 28-72. 1996.

[28] Anders, André, Ian G. Brown, Robert A. MacGill, and Michael R. Dickinson. "Triggerless' triggering of vacuum arcs." Journal of Physics D: Applied Physics 31, no. 5 (1998): 584.

[29] Teel, George, Alexey Shashurin, Xiuqi Fang, and Michael Keidar. "Discharge ignition in the micro-cathode arc thruster." Journal of Applied Physics 121, no. 2 (2017): 023303.

[30] Miller, H. Craig. "Flashover of insulators in vacuum: review of the phenomena and techniques to improved holdoff voltage." IEEE transactions on electrical insulation 28, no. 4 (1993): 512-527.

[31] Miller, H. Craig. "Flashover of insulators in vacuum: the last twenty years." IEEE Transactions on Dielectrics and Electrical Insulation 22, no. 6 (2015): 3641-3657.

[32] Gleichauf, Paul H. "Electrical breakdown over insulators in high vacuum." Journal of Applied Physics 22, no. 5 (1951): 535-541.

[33] Neuber, Andreas A., M. Butcher, H. Krompholz, Lynn L. Hatfield, and Magne Kristiansen. "The role of outgassing in surface flashover under vacuum." IEEE Transactions on Plasma Science 28, no. 5 (2000): 1593-1598.

[34] Neuber, A., M. Butcher, L. L. Hatfield, and H. Krompholz. "Electric current in dc surface flashover in vacuum." Journal of applied physics 85, no. 6 (1999): 3084-3091.

[35] Anderson, R. A., and J. P. Brainard. "Mechanism of pulsed surface flashover involving electron-stimulated desorption." Journal of Applied Physics 51, no. 3 (1980): 1414-1421.

[36] De Tourreil, C. H., and K. D. Srivastava. "Mechanism of surface charging of high-voltage insulators in vacuum." IEEE Transactions on Electrical Insulation 1 (1973): 17-21.

[37] Watson, Alan. "Pulsed flashover in vacuum." Journal of Applied Physics 38, no. 5 (1967): 2019-2023.

[38] Masten, G., T. Müller, F. Hegeler, H. Krompholz, L. L. Hatfield, and M. Kristiansen. "Outgassing and plasma development in the early phase of dielectric surface flashover in

vacuum." In High-Power Particle Beams, 1994 10th International Conference on, vol. 1, pp. 335-338. IET, 1994.

[39] Milton, Osboren. "Pulsed flashover of insulators in vacuum." IEEE Transactions on Electrical Insulation 1 (1972): 9-15.